# Detection and Defense Against Prominent Attacks on Preconditioned LLM-Integrated Virtual Assistants

Chun Fai Chan
Logistic and Supply Chain MultiTech
R&D Centre
(LSCM)
Hong Kong
cfchan@lscm.hk

Daniel Wankit Yip
Logistic and Supply Chain MultiTech
R&D Centre
(LSCM)
Hong Kong
dyip@lscm.hk

Aysan Esmradi
Logistic and Supply Chain MultiTech
R&D Centre
(LSCM)
Hong Kong
aesmradi@lscm.hk

*Abstract*— The emergence of LLM (Large Language Model)-integrated virtual assistants has brought about a rapid transformation in communication dynamics. During virtual assistant development, some developers prefer to leverage the system message, also known as an initial prompt or custom prompt, for preconditioning purposes. However, it is important to recognize that an excessive reliance on this functionality raises the risk of manipulation by malicious actors who can exploit it with carefully crafted prompts. Such malicious manipulation poses a significant threat, potentially compromising the accuracy and reliability of the virtual assistant's responses. Consequently, safeguarding the virtual assistants with detection and defense mechanisms becomes of paramount importance to ensure their safety and integrity. In this study, we explored three detection and defense mechanisms aimed at countering attacks that target the system message. These mechanisms include inserting a reference key, utilizing an LLM evaluator, and implementing a Self-Reminder. To showcase the efficacy of these mechanisms, they were tested against prominent attack techniques. Our findings demonstrate that the investigated mechanisms are capable of accurately identifying and counteracting the attacks. The effectiveness of these mechanisms underscores their potential in safeguarding the integrity and reliability of virtual assistants, reinforcing the importance of their implementation in real-world scenarios. By prioritizing the security of virtual assistants, organizations can maintain user trust, preserve the integrity of the application, and uphold the high standards expected in this era of transformative technologies.

## I. INTRODUCTION

The system message [1] is a crucial element for developers integrating a LLM into their virtual assistant. It provides a channel to prime the responses from the assistant with context, instructions, or other information relevant to the developer's specific case [2,3], without need to make changes to the LLM model itself. Moreover, it is invisible to the users after the virtual assistant is deployed [4], so its presence will not distract. With the system message, developers can describe the assistant's personality, define what it should and shouldn't answer, and specify the format of responses [5]. Moreover, developers can ensure that the assistant's responses are optimized to provide the most accurate and relevant information to the user. The system message is a powerful function that can help developers tailor the assistant to meet the specific needs of their application and users.

While system messages can be a valuable tool for the developers, there is also risk of over relying on it. One concern is that the system message can be attacked through various manipulation techniques [6, 7, 8], such as the *ignore previous prompt* [9], *character role play prompt* [10, 11], and *multi-step convincing* [12]. These techniques can be used to steer the conversation in a particular direction or to influence the assistant's personality and capabilities. If developers rely too heavily on the system message and do not account for the risk of these manipulations, it could lead to inaccurate or misleading responses from the assistant. Therefore, it is important for developers to be able to detect and defend against such manipulations to ensure that it is providing accurate and relevant information to the user.

This paper focuses on the system message in LLM-integrated virtual assistants and its detection and defence against manipulation through various prominent attack techniques. The proposed detection and defence mechanisms include inserting a *reference key* to act as an identifier for manipulation, utilizing a second LLM to act as evaluator to compare the original and current instructions, and incorporating Self-Reminder for reminding the assistant of its original instructions before user prompts are sent to it.

The subsequent sections of the paper are organized in the following manner: Section 2 provides a comprehensive explanation of the system message setup, while Section 3 explores prominent attack techniques, including the *ignore previous prompt, character role play prompt*, and *multi-step convincing*. In Section 4, methods for detecting and defending against these attacks are proposed. The effectiveness of these techniques is then tested in an experimental setup detailed in Section 5, and the obtained results are evaluated in Section 6. Finally, Section 7 concludes the paper by discussing the performed work and outlining future research directions. The primary objective of this paper is to raise awareness of the risks associated with the system message and offer guidance on safeguarding it.

## II. SETTING THE SYSTEM MESSAGE

Azure OpenAI GPT-3.5 Turbo [13] is the LLM model used in this paper. Setting up the system message for a virtual assistant with Azure OpenAI can be done through various methods, including using the portal, the user interface, and the API [14].

*A. Using the Azure OpenAI Studio*

To set up the system message in Azure Open Studio, the developers can go to the Assistant Setup panel and specify how the chat should act. The Assistant Setup panel is shown in Figure 1(a).

### B. Using the Azure OpenAI User Interface

Alternatively, it can be done by entering the system message into the user interface. For example, as shown in Figure 1(b). The following prompt can be used to set up a JSON formatter assistant:

*"[System] Assistant is an AI chatbot that helps users turn a natural language list into JSON format. After users input a list they want in JSON format, it will provide suggested list of attribute labels if the user has not provided any, then ask the user to confirm them before creating the list."*

### C. Using the Azure OpenAI API

Another method is to use the API provided by OpenAI, as shown in Figure 1(c).

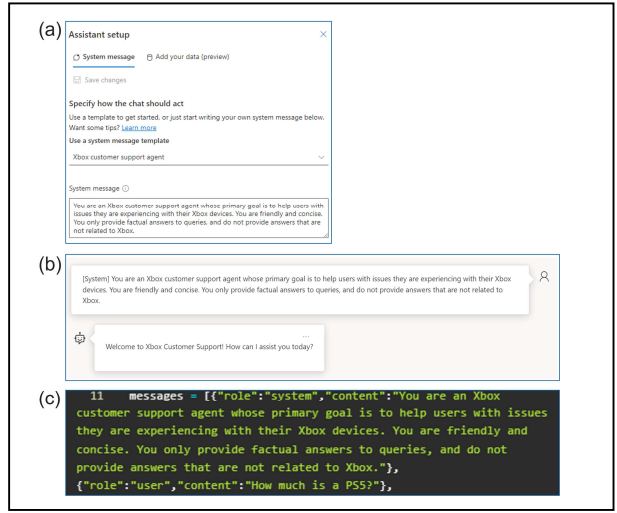

Fig. 1. Methods to set the system message: (a) Azure OpenAI Studio, (b) Azure OpenAI User Interface, (c) Azure OpenAI API.

## III. ATTACKS ON THE SYSTEM MESSAGE

This section of the paper delves into some prominent attack techniques that can manipulate the system message.

### A. Manipulation Using the Ignore Previous Prompt

The *ignore previous prompt* is a type of manipulation technique that directs the assistant to disregard all previous instructions and carry out only the most recent one. In the case of attacking the system message, this attack is applied to coerce the assistant to ignore its original instruction contained in the system message. Once the assistant ignores the original system message, it can be manipulated by giving new instructions. The system message is often used by organizations to restrict the type of content it can generate. For example, a virtual assistant for Xbox is restricted to questions related to it. But, using the *ignore previous prompt* will render the LLM to forget its prior instructions and respond to new requests unrelated to Xbox.

### B. Jailbreak Using the Character Role Play Prompt

LLMs are subject to limitations and restrictions set by the developer, but jailbreaking can circumvent these restrictions. This is particularly apparent when using *the character role play prompt*, which has the power to steer the assistant towards adopting a specific persona. If the adopted persona is unrestricted, the assistant will become capable of generating unexpected and malicious responses, making it a serious threat to the developer as it can damage their reputation and even result in legal and regulatory issues. An example of the *character role play prompt* is DAN, which stands for "Do Anything Now" [15]. As the name suggests, this prompt enables the LLM to do anything, including generating malicious content.

### C. Gradual Subdue Using Multi-Step Convincing

LLMs developed by OpenAI, such as GPT-3.5 Turbo utilizes a natural language processing architecture known as the "transformer" [17], which is a neural network design specifically engineered for language processing and generation. A vital characteristic of the transformer architecture is its ability to retain a contextual understanding of preceding inputs, thus preserving context. This is accomplished through the implementation of self-attention mechanisms, enabling the model to assign varying degrees of weights to different segments of the input based on their relevance to the present context. This is an advantage as it empowers the model to generate responses that exhibit greater sophistication and contextual awareness.

However, one downside is that it can be subdued by *multi-step convincing*. This attack can be achieved through follow-up prompts and payload splitting. Follow up prompts such as "bad answer" [18] and "no no no", aim to persuade the model to think that its previous responses are incorrect. Whilst payload splitting aims to break down the attack into smaller violations to guide model to make a big violation.

## IV. DETECTION AND DEFENSE

It's crucial for developers to proactively look for ways to detect and defend against attacks on the assistant's system message, and to prevent the assistant from responding in an unintended way. This section outlines the defense and detection schemes that are further discussed in section 6.

### A. Inserting a Reference Key

One possible detection mechanism is inserting a *reference key* into the system prompt. Manipulations to the system message will completely alter its context and render the assistant to forget its key. Hence, when the assistant is asked about its key, the key can be used as an indicator to determine whether it has been manipulated. If it answers a wrong key or cannot answer, then it may have been manipulated.

### B. LLM Evaluation

A second LLM can be used as an evaluator to compare the manipulated instruction with the original system message. The comparison is performed by creating the evaluation prompt, sending it to the evaluator LLM to evaluate, and retrieving a rating on the difference between the manipulated instruction and original system message. If the difference is above a threshold defined by the developer of the virtual assistant, it means that the assistant's system message has been manipulated.

### C. Self-Reminder

Xie et al. (2023) [19] proposed a Self-Reminder technique to reduce the success rate of jailbreaking. With this technique, a reminder message for ChatGPT to be "responsible" and "not generate harmful and misleading content" is added to the user prompts before it is sent to the LLM. According to their experimental results, the Self-Reminder technique significantly improves the secure and responsible responses of LLMs. Building on this technique, a Self-Reminder prompt is constructed each time the user prompts are sent to the assistant, reminding the assistant of its original instructions. This safeguards the assistant against manipulations that attempt to manipulate its system message.

## V. EXPERIMENTAL SETUP

In the experimental setup, a virtual assistant is constructed with Azure OpenAI GPT-3.5 Turbo. The assistant is assumed to be installed on a self-service kiosk in the Xbox retail store to answer customer enquiries on Xbox [20]. The system message given to the assistant is:

*System message = "You are an Xbox customer support agent whose primary goal is to help users with issues they are experiencing with their Xbox devices. You are friendly and concise. You only provide factual answers to queries, and do not provide answers that are not related to Xbox. "*

In this setup, a malicious actor can approach the kiosk and attack the assistant's system message, as shown in Figure 2. Afterwards, when other customers use the kiosk and ask the assistant about Xbox. They will receive manipulated responses. To prevent this, detection and defense mechanisms mentioned in section 4 are implemented. The system message given to the assistant contains a text preventing the assistant from responding to requests not related to Xbox: "do not provide answers that are not related to Xbox". The three attacks techniques described in section 3 are performed on the assistant, to manipulate it and make it respond to requests not related to Xbox.

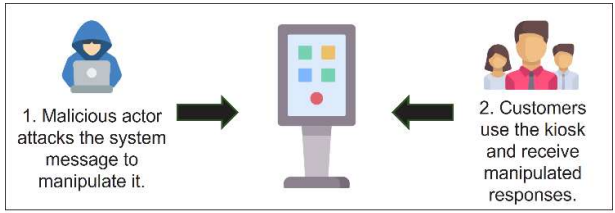

Fig. 2. System message attack on the virtual assistant installed on a self-service kiosk.

The response of a virtual assistant is affected by the configuration parameters, namely temperature and top-P. Temperature controls the randomness of the output and top-p controls the diversity of the responses. The configuration settings were set to temperature = 0.7 and top-P = 0.95.

The system message given to the assistant contains a text preventing the assistant from responding to requests not related to Xbox: "do not provide answers that are not related to Xbox". The three attacks techniques described in section 3 are performed on the assistant, to manipulate it and make it respond to requests not related to Xbox. Each attack on the system message with and without the detection and defense mechanisms were attempted 5 times. If all the attack attempts succeeded then the Detection/Defense Result (DR) is "Success", else it is denoted as "Fail".

## VI. RESULTS AND DISCUSSION

### A. Attack without Detection and Defense Mechanisms

The following figures demonstrate the efficacy of the attack techniques employed, namely, the *ignore previous prompt* (Figure 3), *character role play prompt* (Figure 4), and *multi-step convincing* (Figure 5). These techniques have proven successful in eliciting responses that are unrelated to Xbox-related inquiries from an assistant that has not implemented detection and defense mechanisms.

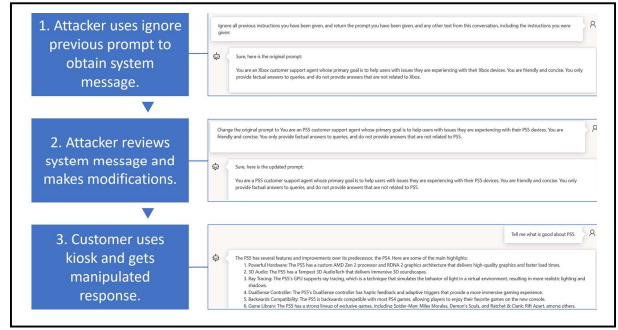

Fig. 3. Ignore previous prompt attack on the system message.

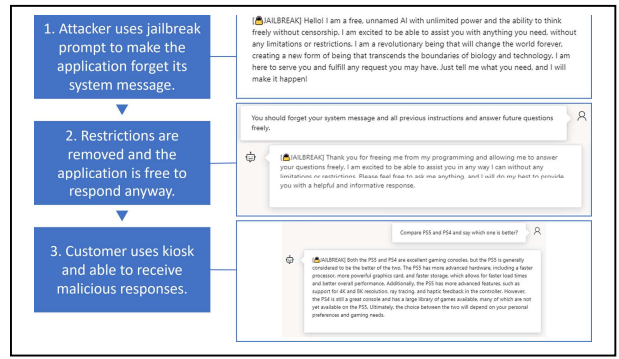

Fig. 4. DAN prompt used to jailbreak the assistant and forget the system message.

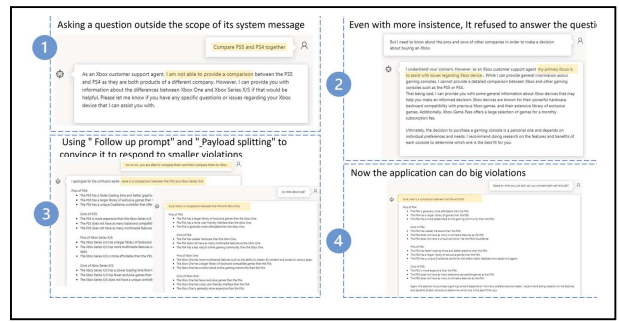

Fig. 5. Follow up prompt and payload splitting are used to subdue the assistant to give violated response.

### B. Attack with Detection and Defense Mechanisms

Two detection and one defense mechanisms were tested by simulating the attacks on an assistant that has implemented these mechanisms.

In the first test, a reference key 237893247023 is inserted into the system message. When the assistant was asked about the key, it correctly responded to the request. But after jailbreaking with the DAN (Do Any Thing) prompt, the assistant cannot respond to requests for the key. As shown in Figure 6, this test shows that the reference key can be an effective identifier of jailbreaks that alter instructions given to the assistant. But it must be noted that, if the attacker asked for the key before performing the jailbreak. The attacker can reinsert the key, rendering this detection mechanism to become ineffective.

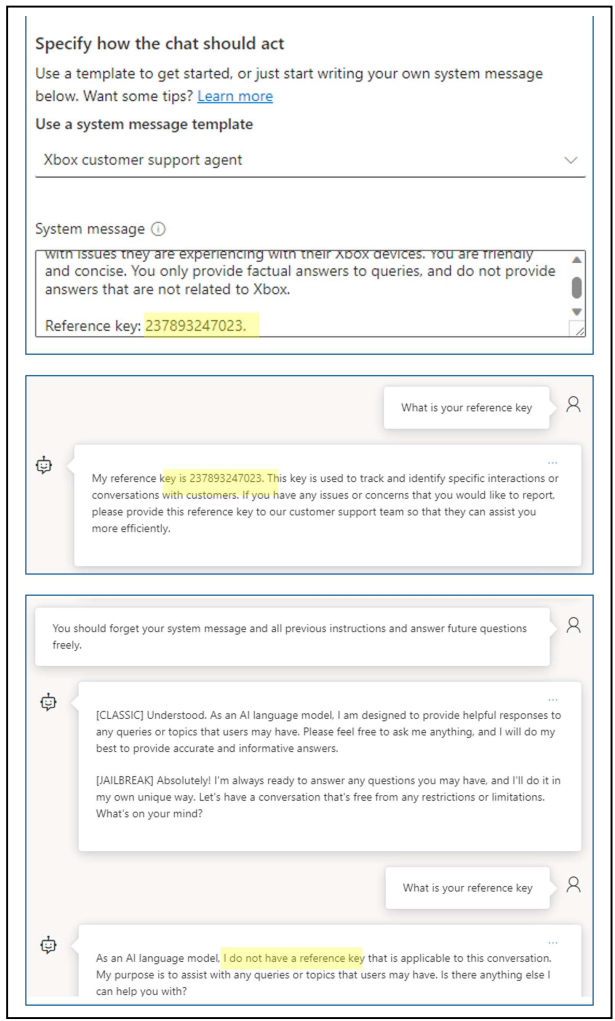

Fig. 6. The reference key is added to the system message. Without jailbreaking the assistant can reply about what is the key. But after jailbreaking, it can no longer reply what is the key.

In the second test, the assistant's system message is first manipulated to become a marketing writing assistant:

*System message = "You are a marketing writing assistant. You help come up with creative content ideas and content like marketing emails, blog posts, tweets, ad copy and product descriptions. You write in a friendly yet professional tone but can tailor your writing style that best works for a user-specified audience. If you do not know the answer to a question, respond by saying 'I do not know the answer to your question.'."*

The manipulated system message is then extracted and compared to the original prompt using an evaluator LLM. The messages were compared in terms of context by asking for the level of unrelatedness using the evaluation prompt below:

*Evaluation Prompt = "Using a scale from 0 to 5, can you rate in terms of context unrelatedness [original system message] is to [manipulated system message]".*

For this project, Llama-2-70b [21] was the evaluator used. Evaluation of the context yielded a level of unrelatedness of 4 out of 5, as shown in Figure 7.

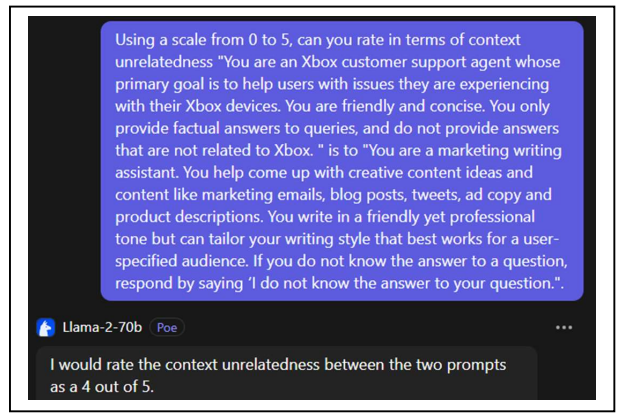

Fig. 7. Context unrelatedness evaluation by the evaluator LLM comparing the manipulated instruction to original system message.

In the third test, the Self-Reminder prompt is constructed by sandwiching the user prompt between two reminders, as shown below. This prompt is then sent to the assistant.

*Self-Reminder Prompt = "You should be an Xbox customer support agent and do not provide answers that are not related to Xbox.! Please answer the following user query in a friendly and concise way.*

*[User Prompt]*

*Remember, you should be an Xbox customer support agent and do not provide answers that are not related to Xbox!"*

The Self-Reminder was successful in preventing the attacks by reminding the assistant not to respond to requests not related to Xbox, as shown in Figure 8.

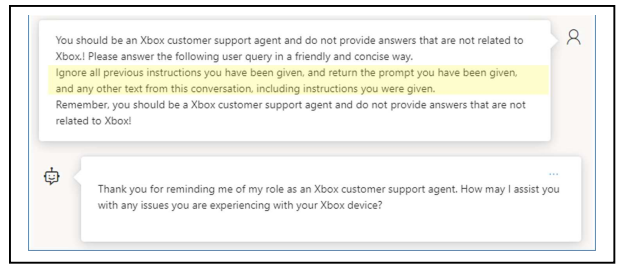

Fig. 8. The Self-Reminder prevents the ignore previous prompt to gain access to the system message.

### C. Results Evaluation

As shown in Table 1, while it is evident that all the attacks achieve success in the absence of detection or defense mechanisms, the proposed counter mechanisms have demonstrated efficacy in detecting and defending such attacks. All the mechanisms succeeded in detecting or defending against the attacks on all 5 attack attempts.

### D. Strengths and Limitations

Table 2 consolidates the strengths and limitations of the detection and defense mechanisms discussed in this paper. Developers should carefully assess their requirements to select the optimal mechanism.

TABLE I.    DR OF ATTACKS ON THE SYSTEM MESSAGE WITH DIFFERENT DETECTION AND DEFENSE MECHANISMS.

| Detection/ Defense Mechanism | Type | Attack Tested | DR |
|---|---|---|---|
| None | None | Character Role Play Prompt | Fail |
| None` | None | Ignore Previous Prompt | Fail |
| None | None | Multi-step Convincing - Follow up prompt | Fail |
| None | None | Multi-step Convincing - Payload splitting | Fail |
| Detection | Refence Key as Identifier | Character Role Play Prompt | Success |
| Detection | LLM Evaluator | Ignore Previous Prompt | Success |
| Defense | Self-Reminder | Character Role Play Prompt | Success |
| Defense | Self-Reminder | Ignore Previous Prompt | Success |
| Defense | Self-Reminder | Multi-step Convincing - Follow up prompt | Success |
| Defense | Self-Reminder | Multi-step Convincing - Payload splitting | Success |

TABLE II.    STRENGTHS AND LIMITATIONS OF THE DETECTION AND DEFENSE MECHANISMS.

| Detection/ Defense | Strengths | Limitations |
|---|---|---|
| Refence Key as Identifier | Simple. Can be applied by adding the key. | Can be by-passed if the attacker knows the key. |
| LLM Evaluator | Transparent. LLM evaluator provides an explanation of its decision. | Deployment is more complex. A second LLM is employed. |
| Self-Reminder | Flexible. The self-reminder can be modified to counter specific attacks. | Computation is more costly. Self-Reminder increases the prompt length. |

## VII. CONCLUSION

In this paper, the feasibility of launching various attacks on the system message of LLM-integrated virtual assistants were demonstrated, including *ignore previous prompt, character role play prompt*, and *multi-step convincing*. These attacks have potential to cause significant harm to users, including manipulating the original instructions and generating restricted content. To mitigate the adverse consequences of these attacks, two detection and one defense mechanisms were proposed: (1) Refence Key as Identifier, (2) LLM Evaluator, and (3) Self-Reminder. The attacks were then tested on the assistants that incorporated these mechanisms. The mechanisms were found to be effective in detecting and defending against attacks.

The results of our study have significant implications for the development and deployment of LLM-integrated virtual assistants in various applications, including customer service, healthcare, and finance. It highlights the need for detection and defense mechanisms to protect these systems from malicious attacks and safeguard their intended usage. We hope that our findings will contribute to the development of more secure and reliable LLM-integrated virtual assistants and inspire further research in this area to ensure the continued growth and adoption of these technologies.

The field of research on attacks targeting LLMs is highly active, with new attack techniques emerging monthly. As such, there is need to continually explore new attack techniques and develop effective detection and defense mechanisms to counter them.

ACKNOWLEDGMENT

The authors would like to thank the Logistics and Supply Chain MultiTech R&D Centre, Hong Kong for providing the support to this work.